# Interaction Between Carbon Nanotubes and Metals: Electronic Properties, Stability, and Sensing

F. Fuchs[a,*], A. Zienert[a], C. Wagner[a], J. Schuster[b], S. E. Schulz[a,b]

[a] Center for Microtechnologies, Reichenhainer Str. 70, Chemnitz, D-09126, Germany
[b] Fraunhofer Institute for Electronic Nano Systems, Technologie-Campus 3, Chemnitz, D-09126, Germany

**Abstract**
The interactions between carbon nanotubes (CNTs) and metal adatoms as well as metal contacts are studied by means of ab initio electronic structure calculations. We show that the electronic properties of a semiconducting (8,4) CNT can be modified by small amounts of Pd adatoms. Such a decoration conserves the piezoelectric properties of the CNT. Besides the electronic influence, the stability of a single adatom, which is of big importance for future technology applications, is investigated as well. We find only small energy barriers for the diffusion of a Pd adatom on the CNT surface. Thus, single Pd adatoms will be mobile at room temperature. Finally we present results for the interaction between a metallic (6,0) CNT and metal surfaces. Binding energies and distances for Al, Cu, Pd, Ag, Pt, and Au are discussed and compared, showing remarkable agreement between the interaction of single metal atoms and metal surfaces with CNTs.

## 1 Introduction

The ongoing miniaturization of electronic devices poses increasing challenges in the field of material science. Some of these demands can be only fulfilled by the introduction of new materials or nanostructures.
Very promising candidates for various applications in micro- and nanoelectronics are carbon nanotubes (CNTs), which offer outstanding electrical [1] and mechanical [2, 3] properties. Depending on their atomistic structure (chirality), there exists a large variety of CNTs with different electronic properties. They can have semiconducting properties, showing a band gap, which is for example useful to produce CNT-based transistors. Metallic CNTs without a band gap can be used in interconnect systems. There are also CNTs with a very tiny band gap, typically below 0.1 eV [4, 5]. They have almost metallic properties and are called semimetallic CNTs. While metallic CNTs do not show any change of the band gap due to uniaxial deformation, the band gap of semiconducting and semimetallic CNTs is sensitive to such a deformation [3, 6, 7, 8]. These types of CNTs are therefore candidates for mechanical sensor devices such as acceleration sensors.
Even though the general physics of CNTs is widely understood, there are still many open questions which are important for technology and which need to be solved on the way towards interconnect and sensor applications of CNTs. One of them is the interaction of the CNTs with metals. Contacts between CNTs and metals are of great importance because metals are used as electrodes in CNT-based circuits. Hereby, a variety of metals with different properties can be used. Finding the most suitable material and understanding the interactions is very important for future devices. In this work, we will present a comparative study of the binding energies between CNTs and the metals Al, Cu, Pd, Pt, Ag, and, Au.

Another important aspect of the interaction between CNTs and metals is the fact, that the decoration of CNTs with metals can be used to adjust the CNT properties [9, 10]. This is interesting as for many applications CNTs with similar properties are required, but the separation of CNTs with desired properties is challenging and expensive. A metal decoration of CNTs can simply be done using e.g. electron beam physical vapor deposition [11]. Thus, a promising idea is to take CNTs which are easy to separate (e.g. a mixture of different types of semiconducting CNTs) and to adjust their properties depending on the application. CNTs could be metalized in order to use them for interconnect systems.

[*] *corresponding author e-mail: florian.fuchs@s2009.tu-chemnitz.de*

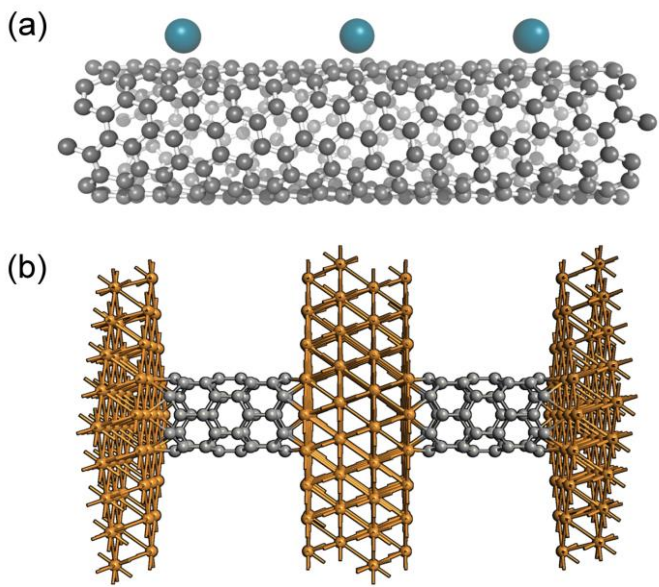

Figure 1: (a) Model of a Pd decorated (8,4) CNT. Three unit cells are shown with one Pd adatom per unit cell. The system is periodic in all three dimensions, but the distance to neighbored CNTs is huge. (b) Model of a (6,0) CNT between flat Cu (111) surfaces, two unit cells are shown. The system is periodic in all three dimensions.

This was already accomplished during our previous studies using small amounts of Co adatoms [10]. A reduction of the band gap is needed for the application of CNTs in sensor devices. For sensors, CNTs with small band gaps are required to obtain sensor currents which are well measurable. Simulations are essential to identify the most suitable metals for the various desired applications and to judge the stability of the binding with the nanotubes. It was shown that particles inside CNTs tend to move along the CNT when an external voltage is applied [12, 13, 14]. A similar behavior is expected for particles or adatoms on a CNT surface. This could strongly reduce the lifetime of devices as, in the worst case, the atoms would simply migrate to one of the contacts. Hence, it is important to study the stability of adsorbed metal on the CNT surface and at the contacts.

## 2 Model system

2.1 Metal adatoms on the CNT surface

For studying metal atoms on a CNT surface we use the semiconducting (8,4) CNT, which has 112 carbon atoms per unit cell and a diameter of about 8.31 nm. We choose this specific tube because it has a small band gap of about 0.74 eV. Furthermore, the diameter of the tube is not unrealistically small, which minimizes possible curvature effects. Nevertheless, it has a manageable system size for complex ab initio studies. All calculations are performed for a single CNT unit cell. We set periodic boundary conditions in all three spatial dimensions and therefore investigate a CNT of infinite length with a selected number of adatoms per unit cell. The lateral extent of the box-shaped unit cell is set to 24.58 Å, so that interactions between periodic neighbors of CNTs are negligible.

The metal atom is added on the CNT surface above the bond of two neighboring carbon atoms, which is the most stable position according to [17]. It will be shown later, that there are only small differences compared to other positions and how an energetically even more favorable position is found. The differences on the electronic properties between these energetically very similar positions are below 0.04 eV and are therefore negligible for judging the electronic properties. For simplicity a uniform and regular decoration of the metal atoms on top of the CNT is considered in the present study. This leads to a chainlike orientation of the adatoms (only one exception will be discussed later, see chapter 4.1 for details). Three unit cells of such a setup are shown in Fig. 1a.

Our study is focused on Pd atoms as decoration metal because it is already used in technology to create low-Ohmic contacts [15] and shows excellent wetting properties on CNTs [16]. Finally, it is a well suited candidate for calculations due to the lack of spin effects – unlike many other metals – when added on CNT surfaces [17]. Spin unpolarized calculations, which are computationally cheaper, are therefore sufficient to get physically correct results.

2.2 CNT-metal contact

The model system used for studying CNT metal contacts consists of a metallic (6,0) CNT which is sandwiched between two flat metal (111) surfaces. So called end contacts are studied, meaning that the CNT is not embedded into the metal but oriented perpendicular to the metal surface in a way that the tube axis is centered on a surface metal atom. There is a finite contact distance between the CNT and the metal which is subject to variation. The system is periodic in all directions. Two unit cells of such a system are depicted in Fig. 1b for the case of Cu contacts. The six fcc metals Al, Cu, Pd, Pt, Ag, and, Au are studied, which are all non-ferromagnetic.

| $N_{Pd}$ | $E_G$ [eV] | $R$ [$10^9$ Ω] | $<\Delta E_{gap}>$ [%] | $s$ [eV] |
|---|---|---|---|---|
| 0 | 0.74 (0.77) | 5.60 (9.01) | - | 0.058 |
| 1 | 0.63 (0.65) | 0.68 (0.93) | 9.4 | 0.061 |
| 2 | 0.61 | 0.43 | 16.8 | 0.048 |
| 3 | 0.52 | 0.08 | 33.7 | 0.027 |
| 4 | 0.13 | 0.000042 | 65.0 | 0.055 |

Table 1: Band gap $E_G$, ballistic resistance R, mean band gap reduction $<\Delta E_{gap}>$, slope s for different amounts of Pd atoms $N_{Pd}$. The values in parenthesis correspond to results obtained using plane wave based DFT.

## 3 Simulation Details

For electronic structure simulations of CNTs a vast variety of different simulation approaches are possible [20]. Recently, we published comparative studies of electron transport in CNTs using density functional theory (DFT), extended Hückel theory, and tight-binding [18, 19].

As the number of atoms per unit cell is relatively small for the topic discussed here, density functional theory (DFT) is a well-suited method for our calculations. The simulations of the metal decoration are performed using the SIESTA package [21], which is based on local atomic orbitals and pseudopotentials. A double zeta polarized basis set was used. The number of k-points along the tube was set to 20 and the mesh cutoff for solving the Poisson equation was set to 100 Ha.

To verify our results, we use the plane wave based DFT code Quantum Espresso (QE) [22] with ultrasoft pseudopotentials. The kinetic energy cutoff of the wavefunctions and for charge density/potential was set to 15 Ha and 120 Ha respectively. Due to the different basis, this method is suitable to verify our results.

While the structures are optimized by SIESTA using the conjugate gradient (CG) method with maximum forces of 0.02 eV/Ang, the Quantum Espresso optimizations are based on the Broyden-Fletcher-Goldfarb-Shanno algorithm with maximum forces of 0.01 eV/Ang. The presented results on metal adatom decoration, unless otherwise indicated, are calculated with SIESTA.

For the results corresponding to the CNT metal contacts, the software package Atomistix ToolKit [23, 24] was used, which is as SIESTA based on local atomic orbitals and pseudopotentials. In this case, the number of k-points and the mesh cutoff was set to 1 (since the unit cell is large) in each spatial region and 75 Ha respectively.

All calculations used for this work are performed using the GGA functional of Perdew, Burke, and Ernzerhof [25].

## 4 Results and Discussion

4.1 Properties of Pd decorated CNTs

To judge the electronic properties of the system, we calculate the band structure (Fig. 2) for all configurations of Pd adatoms and extract the band gap $E_G$. The band gap of a pristine (8,4) CNT is at about 0.74 eV. This value is in good agreement with comparable DFT calculations [26], but smaller than experimental results (where values above 1.1 eV were found [27, 28]). However, as the influence of the metal on the electronic properties of the tube is the main focus of this work, the change of the band gap is more important than its actual value, so the DFT value is sufficient.

By adding one single Pd atom on the surface, the band structure is changed and the valence and conduction bands move slightly to the Fermi level (Fig. 2). Hereby, the band gap is reduced to 0.63 eV, which corresponds to a reduction of about 15 %. Using plane wave based DFT, we obtain values of 0.77 eV and 0.65 eV for the pristine and decorated CNT respectively. The differences between both methods are small and the reduction of the gap due to the metal decoration is in good agreement.

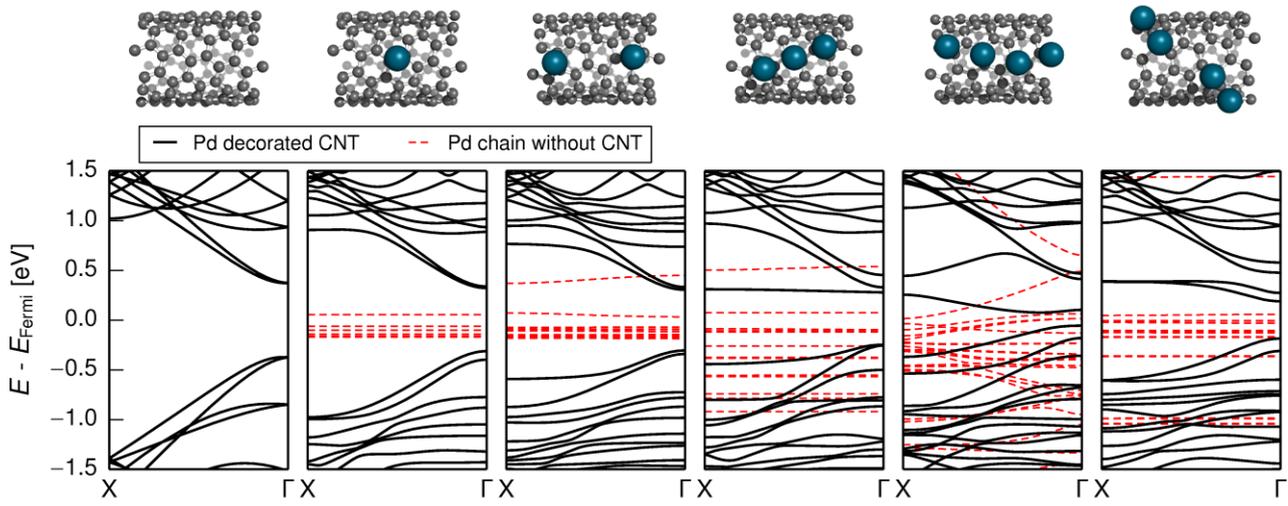

Figure 2: Band structures of CNTs with different amounts of Pd adatoms (black, solid lines). The structures on top correspond to the results shown directly below. In addition, the band structures of the respective Pd chains without CNT are shown (red, dashed lines). For four Pd atoms, two different configurations are shown. See text for details.

The band structure can be manipulated even further by increasing the number of Pd atoms (see Fig 2). The corresponding values of the band gap are shown in Fig. 3 (the numerical values are given in Tab. 1). Up to three Pd adatoms the band gap scales nearly linearly with the number of adatoms. However, four Pd atoms on the CNT reduce the band gap drastically to 0.13 eV (82 % reduction compared to the pristine CNT). In order to understand this behavior, we calculate the electronic properties of the bare chain of Pd atoms (Fig. 2). For one and two Pd atoms, the flat bands indicate almost no interaction between the atoms. In case of three atoms the bands split apart due to orbital overlap, but still no charge transport can occur because the Pd atoms on the CNT surface form very short chains (consisting of three atoms each), which are almost isolated from each other. This is different for the case of four Pd atoms. Here, a direct charge transport through the adatoms is possible as there is a closed metal chain along the whole CNT, which is indicated by the strong bending of the bands. This causes the observed reduction of the band gap and resistance of the total system. In order to check this finding, we calculate a different four-atom configuration, where the four atoms are displaced in such a way that the chain is destroyed (the last configuration in Fig. 2) and direct transport through the metal is expected to vanish. This configuration is less stable as the chain-like configuration (the total energy difference between both configurations with four atoms was about 1.33 eV in favor of the closed Pd chain). Indeed, in the case of the destroyed chain we find a lower value of the gap reduction for the four atom configuration which fits also well to the previously observed linear scaling of the band gap reduction with the number of adatoms (see Fig. 3). The ballistic resistance $R$ of CNTs can be calculated approximately based on a simple formula [2]

$$R(E_G) = \frac{1}{8} h e^{-2} \left[ 1 + \exp\left(\frac{E_G e}{2 k_B T}\right) \right] . \qquad (1)$$

The results for the Pd decorated CNTs are shown in Tab. 1. The case of four Pd atoms is of special interest. The resulting resistance (0.042 µΩ) is very close to the resistance of semimetallic CNTs (typically below 0.03 µΩ), meaning that the electronic properties of semiconducting CNTs can be successfully transformed to resemble semimetallic ones.

Sensing of deformations due to vibration or acceleration is a very promising application where CNT-based sensors are expected to outperform conventional sensors due to their outstanding piezoresistive properties [3, 6, 7, 8]. Thus it is an important question whether the piezoresistive behavior of the (8,4) CNT is conserved after metal decoration. Hence, we stepwise deform our system in uniaxial direction in order to investigate the strain sensing behavior. Fig. 4 shows the dependence of the band gap and the ballistic resistance on the deformation ε for the pristine and single Pd decorated CNT. Results for both DFT methods are shown. Even though there is a small shift of the data between both methods, a linear behavior of the band gap up to strain values of about 10 % for the pristine as well as the decorated CNT is predicted (by both methods). For higher decoration, the deformation dependence is shown in Fig. 5.

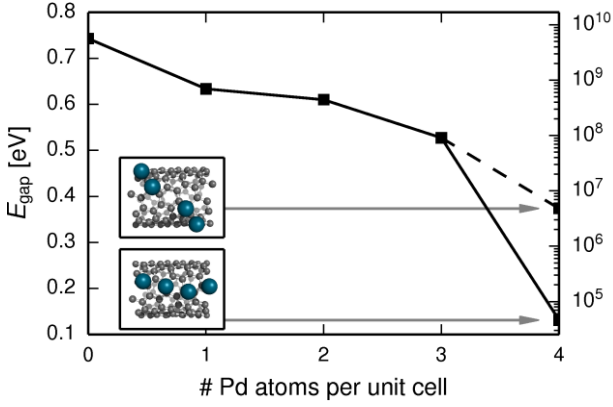 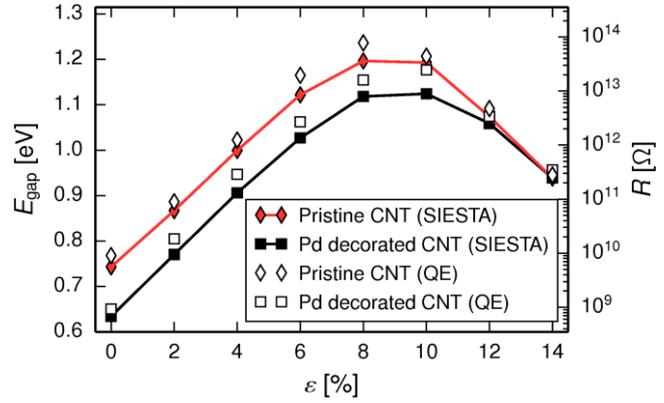

*Figure 3: Band gap vs. number of Pd atoms. For four atoms, two different configurations are shown.*

*Figure 4: Strain dependence of the band gap and the resistance for the pristine and Pd decorated (8,4) CNT, calculated with two different methods.*

To quantify the band gap dependence, we use a simple linear fit procedure of the first five deformation steps. After that, we can calculate the slope s, which is defined as

$$s = \Delta E_{\text{gap}} / \Delta \varepsilon \qquad (2)$$

and the mean band gap reduction $<\Delta E_{\text{gap}}>$. The latter is defined as the mean of the band gap differences between decorated and pristine CNT, where again the first five deformation values were considered for the average. These quantities are listed in Tab. 1. We find very similar slopes for the investigated decoration rates. One exception is the case of three atoms. Here, the adatoms have the tendency to form a short metal chain on the CNT, which reduces the strain sensibility of the CNT. In all other cases, the sensing behavior of the underlying CNT is conserved. This is of big importance as it shows that Pd decorated CNTs can be used in future strain sensors.

In conclusion, the Pd decoration of the semiconducting (8,4) CNT shifts the band gap of the CNT towards values which are comparable to semimetallic CNTs while the sensitivity to mechanical deformation is conserved. This enables sensor applications using Pd decorated semiconducting CNTs. In addition, metal decoration allows for a systematic tuning of the band gap which decreases linearly with the Pd coverage.

It must be noted that clustering of metal adatoms was observed in experiments with various metals such as Au or Al [29, 30, 31]. For Pd, both clustering [11, 31] as well as continuous coating has been achieved [29]. To explain these experimental findings, further investigations of other Pd atom configurations are necessary to find the energetically most favorable geometries (e.g. tetrahedral arrangements). However, since the goal of the present work is to demonstrate the influence of a few Pd atoms per unit cell on the electronic properties of the CNT, our research concentrates on the chain-like structures and other configurations will be subject of future studies.

4.2 Binding stability of single Pd adatoms

For the purpose of binding stability investigations, we calculate an energy landscape for a single Pd adatom on the CNT surface. For the calculation of the energy landscape, the Pd atom is added on different positions inside a rectangular area on the CNT surface. A relaxation algorithm (CG method) is applied, where we constrain the Pd atom to move only perpendicular to the CNT surface, and only nearby carbon atoms (within a range of 4 Å from the Pd position) are allowed to rearrange, while the others are fixed. In contrast, a full energy minimization would not give information about regions outside energetic minima. The described procedure provides an energy value for every position of a single Pd adatom on the CNT surface. The resulting energy landscape is shown in Fig. 6a. It can be seen that there is a metastable position in the middle of a hexagon. Local minima between carbon hexagons in the direction of the tube are visible. However, a valley of small energy values along the CNT is of largest importance. Fig. 6b shows the energy along that valley. We find energy barriers of about 0.04 eV for the motion of a Pd adatom along the energetic valley. This is comparable to the thermal energy at room temperature (0.025 eV), which means that a diffusion of adatoms along the

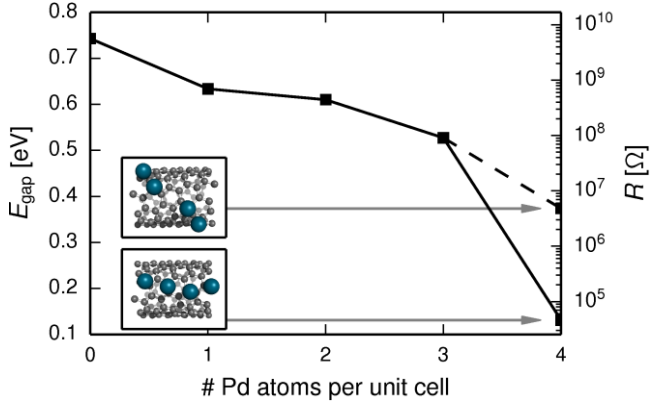 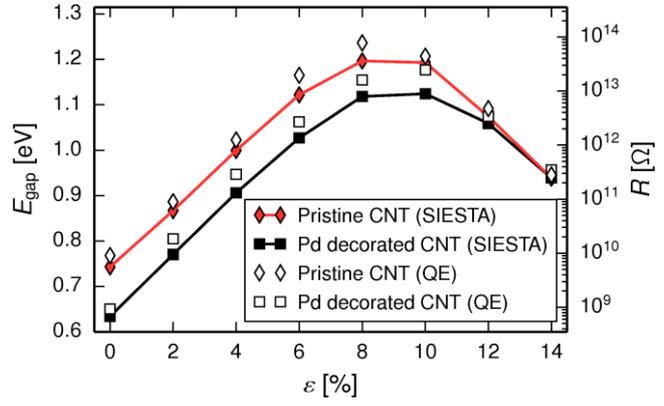

Figure 5: Strain dependence of the band gap and the resistance for the pristine and Pd decorated (8,4) CNT with different numbers of Pd adatoms per unit cell. The solid lines are obtained by a linear fit.

Figure 6: (a) Total energy of the Pd decorated system as a function of the Pd adatom position on the CNT sidewall ($x_{surface}$ denotes the position on the CNT surface around the CNT, z the position along the CNT). A valley in the energy landscape is clearly visible. White circles indicate the extracted values for Fig. 6b. Carbon atoms are marked by grey circles. (b) Energy along that valley. Arrows indicate the corresponding positions within the energy landscape. Values are relative to $E_{min} \approx -18.95$ keV. Results of different full relaxations are shown too. See text for details.

CNT surface should be possible (e.g. by thermal activation or by additional forces due to applied voltage).

For comparison, the data points obtained by a full relaxation of the model system with different initial Pd positions are given, too. For most of the starting positions, the Pd atom approaches a local minimum. There are a few exceptions, where the relaxation stops with a Pd atom at a metastable position. At these points, a saddle point can be found in the energy landscape and as the forces acting on the adatom are low near a saddle point, the optimization algorithm stops as soon as the convergence criterion is reached. The results of the full relaxation are in a very good agreement with the values of the energy landscape, justifying the method explained above.

In summary, there are only small energy barriers for the motion of a single Pd atom on an ideal CNT surface. This is in good agreement with previous simulations of Pd decorated graphene, where also only small energetic differences between different Pd positions were found [31]. However, as metals have the tendency to form clusters, the stability of such clusters on the CNT surface needs to be investigated in the future. Metal clusters are expected to show a higher stability due to the increased contact area. Finally, the presence of defects in realistic CNTs offers very stable positions for the decoration, as was recently found for Pd decorated vacancy defects in graphene [31, 32]. The investigation of the electronical and electromechanical influence of Pd atoms on defects will also be subject of upcoming work.

4.3 Interaction between CNTs and metal surfaces

Besides using metal atoms for tuning the electronic properties of CNTs, metals are often used as contact materials in CNT-based devices. It is known that different applications (e.g. sensors, interconnects, or transistors) require different types of contact (e.g. Ohmic or Schottky barrier). Besides its electronic properties, a high mechanical stability of the contact is always desirable. Based on the model system shown in Fig. 2b, we analyzed the stability of CNT end contacts to different fcc metals. There are 6 carbon atoms with unpassivated bonds at the CNT-contact interface. Such a configuration differs from experimental setups, where the CNT is often embedded. However, a similar trend regarding the potential of the metals to form stable contacts is expected. Fig. 7a shows the total energy of the model system as a function of the CNT metal contact distance for various metals. The continuous curves are the result of a simple double exponential fit

$$E(d) = A(\exp(-B_1 d) - 0.25\exp(-B_2 d)) , \qquad (3)$$

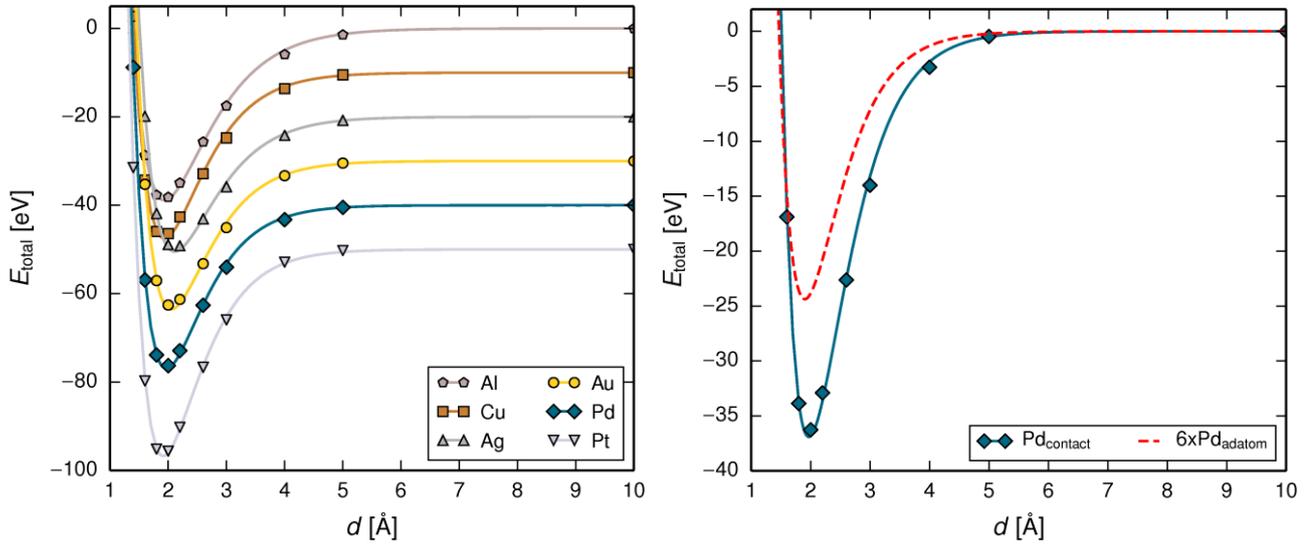

Figure 7: Left: Total energies versus binding distance for different metal surfaces, where the curves are shifted for clarity. The lines are obtained by a fitting procedure. Right: Same as in the left picture, but for the two different model systems using Pd discussed here. The values for the decoration are multiplied by 6 and the curve is shifted so that the two saturation points overlay.

where $A$, $B_1$ and $B_2$ are fit parameters. The fitted curves reveal the binding energy between the CNT and the metal surface and the equilibrium distance can be derived (Tab. 2). Based on these data, the metals can be ranked with respect to their binding strength against flat metal surfaces:

$$\underbrace{E_b(\text{Ag})}_{82\%} < \underbrace{E_b(\text{Au})}_{91\%} < \underbrace{E_b(\text{Pd})}_{100\%} < \underbrace{E_b(\text{Cu})}_{102\%} < \underbrace{E_b(\text{Al})}_{106\%} < \underbrace{E_b(\text{Pt})}_{128\%}. \qquad (4)$$

The noble metals Ag and Au form similarly weak contacts, while Cu, Pd, and Al are intermediate. Pt forms the most stable contacts. A comparative study of the metals regarding their ability to form good electrical contacts to CNTs can be found in [33].

It is interesting, to compare the binding properties between single Pd adatoms on a CNT surface and CNTs in end-to-end contacts with a flat Pd (111) surface. The distance dependence in both cases is remarkably similar, as shown in Fig. 7b. Apart from the different binding energy, due to the different number of bonds, both the binding distance and the shape of the $E(d)$ curves are in reasonable agreement. This is surprising as both structures are very different and dangling bonds are present at the interface in case of the end-to-end contact. We expect that this holds for other metals, too.

## 5 Conclusions

We could demonstrate that the electronic properties of a pristine (8,4) CNT can be adjusted by a decoration with Pd adatoms. In case the Pd atoms create a closed chain on the CNT surface, the band gap is drastically reduced, leading to almost similar electronic behavior compared to semimetallic CNTs.

Furthermore, the strain sensitivity of the band gap with respect to the deformation of the underlying CNT is conserved for the observed configurations. This means, that such a CNT can be used to produce strain sensor devices.

However, the stability of a single Pd adatom is relatively small on the CNT surface, which can lead to diffusion of the metal atoms along the CNT. The stability of metal clusters and the role of defects will be the issue of future investigations.

The investigation of the binding abilities between CNT and metal surfaces allows a ranking of the investigated metals with respect to the binding energies. We could show that Pt forms the most stable contacts, followed by Al, Cu, and Pd. Contacts between the metals Au and Ag and CNTs are least stable.

**Acknowledgements**
We acknowledge the ongoing support by the group of Michael Schreiber. This work was funded by the DFG Research Unit 1713 "Sensoric Micro- and Nano- Systems" (SMINT) and by the International

| Metal | $d_{opt}$ [Å] | $E_{bond}$ [eV] |
|-------|---------------|-----------------|
| Al    | 1.914         | 19.3            |
| Cu    | 1.915         | 18.7            |
| Pd    | 1.960         | 18.3            |
| Pt    | 1.908         | 23.4            |
| Ag    | 2.113         | 15.1            |
| Au    | 2.048         | 16.7            |

Table 2: Optimal distances $d_{opt}$ and total binding energy $E_{bond}$ between CNT and different metal surfaces.